\documentstyle[12pt]{article}
\def\bc{\begin{center}}
\def\ec{\end{center}}
\def\be{\begin{equation}}
\def\ee{\end{equation}}
\def\bea{\begin{eqnarray}}
\def\eea{\end{eqnarray}}

\def\simge{\ \lower-
1.2pt\vbox{\hbox{\rlap{$>$}\lower5pt
\vbox{\hbox{$\sim$}}}}\ }
\def\cs{\left. \mu {d \over{d \mu}} \right|_{g_0,a}}
\def\csr{(\mu {{\partial} \over {\partial \mu}} +\beta(g)
{{\partial}\over {\partial g}})}
\def\csm{\left. \mu {d \over{d \mu}} \right|_{g_0,M_0,a}}
\begin{document}
\pagestyle{empty} 
\vspace{-0.6in}
\begin{flushright}
\end{flushright}
\vskip 2.0in
\centerline{\large {\bf{SOME OBSERVATIONS ON BROKEN SYMMETRIES}}}
\vskip 1.0cm
\centerline{M. Testa$^{1,2}$}
\centerline{\small $^1$  Theory Division, 
CERN, 1211 Geneva 23,
Switzerland$^{\star}$.}
\centerline{\small $^2$ Dipartimento di 
Fisica, Universit\`a di Roma ``La
Sapienza"}
\centerline{\small Sezione INFN di Roma}
\centerline{\small P.le A. Moro 2, 00185 
Roma, Italy$^{\star}$$^{\star}$.}
\vskip 1.0in
\abstract{We present a general analysis of the field theoretical
properties which guarantee the recovery, at the renormalized level,
of symmetries broken by regularization.
We also discuss the anomalous case.}
\vskip 1.0in
\begin{flushleft} 
\end{flushleft}
\vfill
\noindent \underline{\hspace{2in}}\\
$^{\star}$ Address until August 31st, 1998.

\noindent $^{\star}$$^{\star}$ Permanent address.
\eject
\pagestyle{empty}\clearpage
\setcounter{page}{1}
\pagestyle{plain}
\newpage 
\pagestyle{plain} \setcounter{page}{1}
\section{Introduction}

The situation in which a classical global symmetry is broken
through regularization is far from exceptional in field theory
and is relevant for
basic symmetries such as chirality and scale invariance.
In these cases either the symmetry is restored in the infinite cutoff
renormalized theory or quantum anomalies definitively spoil the
conservation laws of the classical action.

While for global symmetries respected by the regulator
the bare Noether currents are finite and correctly normalized,
the situation is different for classical symmetries
violated by the regularization. In this case, in fact,
even if the symmetry will be recovered in the renormalized theory,
the corresponding currents require finite renormalization.
In the presence of anomalies, instead, infinite, scale dependent
renormalizations are needed in order to define the corresponding currents.

The general requisites underlying the recovery of the non-anomalous
$SU(3)\bigotimes SU(3)$ chiral symmetry and the corresponding
classification of the local observables, have been discussed in ref.\cite{boc}
in the context of the lattice discretization with Wilson
fermions\footnote{As repeatedly stressed, the discussion presented in
ref.\cite{boc} is not restricted to lattice regularization,
but is indeed quite general and it is relevant
whenever the process of regularization breaks a symmetry of
the classical action.}.
In the case of octet chiral currents, some of these requirements,
those underlying the construction of the conserved currents,
have been shown\cite{cur} to be satisfied at any order in
perturbation theory. 
It must be said, however, that checks within perturbation theory,
although important, do not provide an "explanation" why these
prerequisites are indeed satisfied: one is left with the impression
that the theory is, for some mysterious reason, ready to
incorporate the symmetry, although broken for any finite cutoff.

It is the purpose of this paper to explore the mechanisms
underlying the recovery of a symmetry, without invoking
perturbation theory, but only using general properties expected
to be non-perturbatively valid in any field theory.
In other words, we will try to trace back to some general field theoretical
features the various properties which, in ref.\cite{boc}, were conjectured
from the analysis of one loop perturbation theory.

Let me state explicitly the general euclidean field theoretical
assumptions which I will use (at least those of which I am aware):

\begin{enumerate}

\item the renormalization structure is the same as
the one found in perturbation theory; \label{first}

\item renormalized Green's function are finite when computed
at non-exceptional external momenta and obey the Callan-Symanzik
equation\cite{cal}; \label{third};

\item the theory exists in the zero mass limit,
in the sense that Green's functions of "suitably
renormalized operators"\footnote{This
means that the renormalization conditions should not
introduce spurious, mass-dependent i.r. divergent terms like,
e.g. $log(p^2/m_q^2)$.
The i.r. safe renormalization conditions are the usual ones performed
at a scale $\mu$.} are finite when computed at
non-exceptional momenta\cite{poggio}; \label{fourth}

\item the \emph{regularized} Green's functions are finite
at non-exceptional momenta, also in the massless limit \label{erre}.
\end{enumerate}
 
The above properties are known to be true to all orders in perturbation
theory. If they will ever turn out to have a non-perturbative
validity, the same will be true for the considerations exposed in the following
sections.
Assumption \ref{erre} is, however, on a different ground:
it concerns the regularization.
Although valid for any decent regularization, we will show
that it is not necessary in the proof of
the recovery of the symmetry, even in the massless limit:
assumption \ref{erre} will only be used to exclude some pathological
behaviour, not found in lattice perturbation theory, of the renormalization
constants of \emph{any} operator.

The plan of the paper is as follows.
In Section \ref{one} we give a simple non-perturbative derivation
of a fundamental, well-known, order by order in perturbation
theory, property concerning composite operators:
power divergent mixings with lower dimensional operators
never depend on the renormalization scale $\mu$.
Starting from this result we will discuss, in Section \ref{two},
the case of Ward Identities related to global non-anomalous symmetries,
showing their validity in the renormalized theory.
Finally, in Section \ref{three}, we will discuss the case of the chiral
$U(1)$ global anomalous symmetry and its Ward Identities.

\section{Power Divergent Operators}\label{one}

We start considering a theory with fundamental fields $\phi(x)$,
regularized by a cutoff\footnote{We will, for definiteness, adopt
the language of lattice regularization,
but, as already stressed, our discussion is completely general.}
$\Lambda =1/a$,
defined by a certain number of (bare) coupling constants which we
collectively denote by $g_0$. The theory is renormalized at a subtraction
point $\mu$ and the renormalized Green's functions, $G^{(n)}(x;g,\mu)$,
are expressed in terms of a set of renormalized couplings $g$:
\be
G^{(n)}(x;g,\mu) \equiv {1 \over {[Z_\phi(g_0,a\mu)]^{n/2}}}
\langle \phi(x_1) \dots \phi(x_n) \rangle \equiv 
\langle \phi^{(ren)}(x_1) \dots \phi^{(ren)}(x_n) \rangle \label{ren}
\ee
where $Z_\phi(g_0,a\mu)$ is the wave function renormalization
of $\phi(x)$, $\phi^{(ren)}(x)$ denote the renormalized fields and
$\langle \dots \rangle$ is the euclidean expectation value with
respect to the regularized measure.

Just to establish some notation, let us recall that the
Callan-Symanzik\cite{cal} differential operator,
$\cs$ acting on
$G^{(n)}(x;g,\mu)$ gives:
\begin{eqnarray}
& & \cs G^{(n)}(x;g,\mu) \equiv
\csr G^{(n)}(x;g,\mu)= \label{cal} \\
& & = \left[\cs {1 \over {[Z_\phi(g_0,a \mu)]^{n/2}}} \right]
\langle \phi(x_1) \dots \phi(x_n) \rangle = \nonumber \\
& & = -n \gamma_\phi(g) G^{(n)}(x;g,\mu) \nonumber
\end{eqnarray}
where, as usual, we denoted by $\gamma_\phi(g)\equiv {1 \over 2}
\left. \mu {d \over{d \mu}} \right|_{g_0,a} \log [Z_\phi(g_0,a \mu)]$
the anomalous dimension of $\phi$. In eq.(\ref{cal}) we exploited the
fact that $\cs$
gives $0$, when acting on bare quantities.

Let us now consider the case of a composite operator $O(x)$. In order to make
it finite it has to be mixed with bare operators of equal or smaller
dimension. In order to simplify the presentation of the
argument we consider a simple situation
in which only mixings with lower dimensional operators occur.

In this case we have:
\begin{equation}
O_R(x)=Z_O [O(x)+{\tilde Z \over a} \tilde O(x)] \label{comp}
\end{equation}
where we schematically denoted by $\tilde O(x)$ the lower dimensional
operators.

The dimensionless coefficients $Z_O$ and $\tilde Z$ are chosen,
according to
appropriate renormalization conditions at the scale $\mu$, so
that the Green's functions:
\begin{equation}
G^{(O_R,n)}(x, x_1 \dots x_n) \equiv {1 \over {[Z_\phi(g_0,a \mu)]^{n/2}}}
\langle O_R(x) \phi(x_1) \dots \phi(x_n) \rangle \label{greencomp}
\end{equation}
stay finite, together with their Fourier transforms, as $a \rightarrow 0$.

Applying $\cs$ to
both sides of eq.(\ref{greencomp}), we get:
\begin{eqnarray}
& & \cs
G^{(O_R,n)}(x, x_1 \dots x_n) \equiv \label{csop} \\
& & \equiv \csr G^{(O_R,n)}(x, x_1 \dots x_n)= \nonumber\\
& & = -n \gamma_\phi(g) G^{(O_R,n)}(x, x_1 \dots x_n)+ \nonumber \\
& & +{{\left[ \cs Z_O \right]} \over Z_O}
 G^{(O_R,n)}(x, x_1 \dots x_n)+ \nonumber \\
& & + {Z_O \over a} \left[\cs \tilde Z\right]
\langle \tilde O(x) \phi_R(x_1) \dots \phi_R(x_n) \rangle \equiv \nonumber \\
& & \equiv -( n \gamma_\phi(g) + \gamma_O(g))
G^{(O_R,n)}(x, x_1 \dots x_n)+ \nonumber \\
& & +{Z_O \over a} \left[\cs \tilde Z\right]
\langle \tilde O(x) \phi_R(x_1) \dots \phi_R(x_n) \rangle \nonumber
\end{eqnarray}
where $\gamma_O(g)$ denotes the anomalous dimension of $O_R(x)$.

Since the l.h.s. and the first term of the r.h.s. of eq.(\ref{csop})
are finite, we must have:
\begin{equation}
\cs \tilde Z=0 \label{res}
\end{equation}

Eq.(\ref{res}) proves in complete generality that power
divergent subtractions of composite operators do not contribute to
anomalous dimensions and that, in presence
of dimensionless couplings only, they do not contain any logs.

\section{Global Non-Anomalous Symmetries} \label{two}

In this section we will deal, for definiteness, with non singlet
chiral transformations, softly broken by a
mass term\footnote{For simplicity we will consider in the
following the flavor-symmetric case $M_0 \propto I$} $M_0$.

\subsection{Ward Identities with Elementary Operator Insertions}

As discussed in ref.\cite{boc}, the naive chiral variation
of the regularized functional integral
gives rise to the regularized Ward Identity\footnote{We adopt the conventions:
$\{\lambda^a, \lambda^b \} =
2 d^{abc} \lambda^c$, $d^{ab0}= \delta^{ab}$, $\lambda^0 = {2 \over3} I$.}:
\bea
& & \partial^\mu_x \langle A^a_\mu(x) \ q(y) \ \bar q(z) \rangle=
M_0 \langle \bar q(x) \lambda^a \gamma_5 q(x) \ q(y) \ \bar q(z) \rangle
+ \label{nward}\\
& & + \delta^{(4)}(x-y) \langle \gamma_5 {\lambda ^a \over 2} q(y) \
\bar q(z) \rangle
+ \delta^{(4)}(x-z) \langle q(y) \ \bar q(z) {\lambda ^a \over 2} \gamma_5
\rangle + \nonumber \\
& & + \langle X^a(x) \ q(y) \ \bar q(z) \rangle \nonumber
\eea
where 
\be
A^a_\mu(x) \equiv \bar q(x) {\lambda^a \over 2} \gamma_\mu \gamma_5 q(x)
\ee
and $X^a(x)$ is the chiral variation of the Wilson term.
Since $X^a(x)$ vanishes in the formal classical limit, it has the form:
\be
X^a(x)=a O_5^a(x) \label{ordera}
\ee

Eq.(\ref{nward}) has still to be renormalized.
In particular the composite operator $O_5^a(x)$, defined in
eq.(\ref{ordera}), is not finite as $a \rightarrow 0$, but
contains power divergences which can compensate the overall factor $a$.
In order to construct a finite operator, $\bar O_5^a(x)$,
out of $O_5^a(x)$, we must consider appropriate linear combinations,
as generically described in eq.(\ref{comp}). In the case of $O_5^a(x)$,
dimension $5$ operators will appear with logarithmically divergent
coefficients, while lower dimensional operators will contribute with
coefficients proportional to inverse powers of the lattice spacing $a$.
In order to simplify the presentation we will not write down explicitly the
logarithmically divergent mixings. We, therefore, have:
\be
\bar O_5^a(x)= Z_5 \{ O_5^a(x) + {\bar M \over a} \bar q(x)
\lambda^a \gamma_5 q(x)+ {(Z_A -1) \over a}
\partial^\mu
A^a_\mu(x) \} \label{bar}
\ee
While\footnote{Strictly speaking $Z_5$ could also depend on $aM_0$, if we do
not use assumption \ref{erre}. See the further
discussions on this point, later in this section.}
$Z_5(g_0, a \mu)$ is logarithmically divergent, $Z_A$ and
$\bar M$ are restricted by eq.(\ref{res}), to be of the
form\footnote{In this case also $M_0$ has to be included
among the bare couplings.}:
\bea
& & Z_A = Z_A(g_0, a M_0) \label{irz} \\
& & \bar M = {w(g_0, a M_0) \over a} \label{mbar}
\eea
Eq.(\ref{nward}) then becomes\cite{boc}:
\bea
& &\partial^\mu_x \langle \hat A^a_\mu (x) \ q(y) \ \bar q(z) \rangle=
m_q \langle \bar q(x)
\lambda^a \gamma_5 q(x) \ q(y) \ \bar q(z) \rangle + \label{quark} \\
& & + \delta^{(4)}(x-y) \langle \gamma_5 {\lambda ^a \over 2} q(y) \ \bar q(z)
\rangle
+ \delta^{(4)}(x-z) \langle q(y) \ \bar q(z) {\lambda ^a \over 2} \gamma_5
\rangle +\nonumber\\
& & + \langle \bar X^a(x) \ q(y) \ \bar q(z) \rangle \nonumber
\eea
where:
\bea
& & \hat A^a_\mu (x) \equiv Z_A A^a_\mu (x) \nonumber \\
& & m_q \equiv M_0-\bar M \label{chibar} \\
& & \bar X^a(x) \equiv {a \over {Z_5}} \bar O_5^a(x) \nonumber \\
\eea

The $a M_0$ dependence in eqs.(\ref{irz}) and (\ref{mbar}) is not
further restricted by the renormalization group and looks somewhat
problematic, at least for considerations pertaining the massless limit of QCD,
reached when:
\be
M_0=M_{cr} \equiv {f_{cr}(g_0) \over a}
\Leftrightarrow m_q=0 \label{mcr}
\ee
In fact eq.(\ref{irz}), does not forbid the presence, in $Z_A$, 
of terms such as $\log (am_q)$; in ref.\cite{boc} the absence of
such terms was explicitly
assumed and this assumption has afterwards been confirmed in perturbation
theory\cite{cur}. In our case these terms are excluded
by assumption \ref{erre} stated in the introduction, which guarantees
the existence, in the massless limit, of any regularized bare
operator. However, as we will show at the end of this section, this
hypothesis does not seem necessary in order to recover the symmetry, even
in the massless limit. For the moment, therefore, we will proceed without
invoking it.

Using eq.(\ref{quark}) we can now establish, in full generality,
the recovery of chiral symmetry in the continuum limit.
First of all, since the quark fields appear homogeneously in eq.(\ref{quark}),
we can proceed to their renormalization:
\bea
& & \partial^\mu_x \langle \hat A^a_\mu (x) \ q^{(ren)}(y)\
\bar q^{(ren)}(z \rangle=
m_q \langle \bar q(x)
\lambda^a \gamma_5 q(x) \ q^{(ren)}(y)\ \bar q^{(ren)}(z)
\rangle + \label{quarkr} \\
& & + \delta^{(4)}(x-y) \langle \gamma_5 {\lambda ^a \over 2}
q^{(ren)}(y) \ \bar q^{(ren)}(z)
\rangle
+ \delta^{(4)}(x-z) \langle
q^{(ren)}(y) \ \bar q^{(ren)}(z) {\lambda ^a \over 2} \gamma_5
\rangle +\nonumber\\
& & + \langle \bar X^a(x) \
q^{(ren)}(y) \ \bar q^{(ren)}(z) \rangle \nonumber
\eea
where:
\bea
& & q^{(ren)}(x) \equiv {{q(x)} \over {\sqrt {Z_q(g_0,a\mu)}}} \\
& & \bar q^{(ren)}(x) \equiv {{\bar q(x)} \over {\sqrt {Z_q(g_0,a\mu)}}}\nonumber
\eea
In eq.(\ref{quarkr}) we can safely drop the insertion
of $\bar X^a(x)$, since $\bar O_5^a(x)$ is finite when inserted together
with renormalized fundamental fields
and $Z_5$ behaves, at most, logarithmically. We then get:
\bea
& & \partial^\mu_x \langle \hat A^a_\mu (x)
\ q^{(ren)}(y) \ \bar q^{(ren)}(z) \rangle=
m_q \langle \bar q(x)
\lambda^a \gamma_5 q(x) \ q^{(ren)}(y)\ \bar q^{(ren)}(z)
\rangle + \label{quarkr1} \\
& & + \delta^{(4)}(x-y) \langle \gamma_5 {\lambda ^a \over 2}
q^{(ren)}(y) \ \bar q^{(ren)}(z) \rangle
+ \delta^{(4)}(x-z) \langle
q^{(ren)}(y) \ \bar q^{(ren)}(z) {\lambda ^a \over 2} \gamma_5
\rangle \nonumber
\eea

Eq.(\ref{quarkr1}) implies, by a standard argument, the separate
u.v. finiteness of both $m_q \ \bar q(x) \lambda^a \gamma_5 q(x)$
and $\hat A^a_\mu (x)$. In fact, integrating eq.(\ref{quarkr1})
over $x$, we get:
\bea
& & m_q \int d^4 x \langle \bar q(x)
\lambda^a \gamma_5 q(x) \ q^{(ren)}(y) \ \bar q^{(ren)}(z) \rangle 
 =\label{renc}\\
& & =- \left[\langle \gamma_5 {\lambda ^a \over 2} q^{(ren)}(y)
\ \bar q^{(ren)}(z) \rangle
+ \langle q^{(ren)}(y) \ \bar q^{(ren)}(z) {\lambda ^a \over 2} \gamma_5
\rangle \right] \nonumber
\eea
Eq.(\ref{renc}) shows that, in virtue of the (assumed) u.v. finiteness
of the Green's functions of renormalized fields, the integrated mass
insertion is finite. Since there are no
operators of dimension $\le 3$ with identically vanishing $x$
integral, which could mix with $\bar q(x) \lambda^a \gamma_5 q(x)$,
we conclude that the non integrated 
mass insertion in eq.(\ref{quarkr1}) is also finite.
This, in turn, shows that $\partial^\mu_x \langle \hat A^a_\mu (x)
\ q^{(ren)}(y) \ \bar q^{ren}(z) \rangle$ is finite by itself.
Since the symmetry is ungauged, there are no
operators of dimension $\le 3$ with identically vanishing 4-divergence
which could mix with $\hat A^a_\mu (x)$. This shows that $\hat A^a_\mu (x)$
is finite and correctly normalized, since it satisfies the
continuum Ward Identity, eq.(\ref{quarkr1})\cite{boc}.

Let me finally discuss the question of the massless limit.
I want to show that assumptions \ref{first}-\ref{fourth},
alone, imply the existence
of $\hat A^a_\mu(x)$ also in the limit $m_q \rightarrow 0$.
The finiteness of insertions of "suitably renormalized operators"
at non-exceptional momenta, assumption \ref{fourth},
cannot be directly invoked here,
because the normalization of $\hat A^a_\mu(x)$ has not been chosen
"suitably", but has been fixed by the theory itself, through eq.(\ref{chibar}).
However $\hat A^a_\mu(x)$ must be proportional to a "suitably renormalized
operator", possibly through a factor which diverges logarithmically as
$m_q \rightarrow 0$. This means that, if we can show the finiteness
of one particular insertion
of $\hat A^a_\mu(x)$ in the massless limit at non-exceptional
momenta, then the current itself will be well defined.
The argument proceeds as follows. Eq.(\ref{renc}), shows that
$m_q \ \int d^4x \bar q(x) \lambda^a \gamma_5q(x)$ and its limit
for $m_q \rightarrow 0$ can never be infinite, because of the
assumed finiteness of the Green's functions of the renormalized quark fields
appearing in the r.h.s.\footnote{This behaviour is, of course, compatible
with the possibility of spontaneous symmetry breaking.}
This implies that, when inserted at non-zero momentum,
$m_q \ \bar q(x) \lambda^a \gamma_5q(x)$ has to vanish, as
$m_q \rightarrow 0$.
If we now take the Fourier transform of eq.(\ref{quarkr1}), with
respect to $x$, at some momentum $k \neq 0$, for $m_q \rightarrow 0$:
\bea
& & \int d^4x \ e^{-ikx} \ \partial^\mu_x \langle \hat A^a_\mu (x)
\ q^{(ren)}(y) \ \bar q^{(ren)}(z) \rangle= \\
& & =i\ k_\mu \ \int d^4x \ e^{-ikx} \ \langle \hat A^a_\mu (x)
\ q^{(ren)}(y) \ \bar q^{(ren)}(z) \rangle = \nonumber \\
& & = e^{-iky} \ \langle \gamma_5 {\lambda ^a \over 2}
q^{(ren)}(y) \ \bar q^{(ren)}(z) \rangle
+ e^{-ikz} \ \langle
q^{(ren)}(y) \ \bar q^{(ren)}(z) {\lambda ^a \over 2} \gamma_5
\rangle \nonumber
\eea
we see that, again in virtue of the assumed finiteness of the Green's
functions of the renormalized quark fields appearing in the r.h.s.,
the insertion of $\ k_\mu \ \int d^4x \ e^{-ikx} \ \hat A^a_\mu (x)$
is finite and, therefore, so is $\hat A^a_\mu(x)$. 

From now on we will assume that the regularization has been performed so
that assumption \ref{erre} is satisfied. We stress again that the recovery
of chiral symmetry in the massless limit has nothing to do with it:
if assumption \ref{erre} were not fulfilled, then \emph{all}
bare operators would be singular in the massless limit.
As a consequence of assumption \ref{erre}, the $aM_0$ dependence in $Z_A$ can
be safely neglected, for asymptotically small $a$, and $Z_A$
will be a function of $g_0$ alone\cite{boc}.

\subsection{Ward Identities with Composite Operator Insertions}

When studying Ward Identities with composite operator insertions,
another general property was needed in ref.\cite{boc}, which
we will now examine in full generality. It concerns the integrated insertion of
$\bar X^a(y)$ together with a composite local operator.

We will discuss the example of the regularized, integrated octet
($a \neq 0$) axial Ward Identity\cite{boc}:
\bea
& & - m_q Z_p \int d^4 x \langle \bar q(x)
\lambda^a \gamma_5 q(x) \ \bar q(0)
{\lambda^b \over 2} \gamma_5 q(0) \ \Lambda_n (y) \rangle = \label{form} \\
& & = d^{abc} Z_p \langle \bar q(0){\lambda^c \over 2} q(0) \ \Lambda_n (y) \rangle
+ Z_p \langle \bar q(0)
{\lambda^b \over 2} \gamma_5 q(0) \ \delta_A^a \Lambda_n (y) \rangle +\nonumber \\
& & + Z_p \int d^4 x \langle \bar X^a (x)
\ \bar q(0){\lambda^b \over 2} \gamma_5 q(0) \ \Lambda_n (y)\rangle \nonumber
\eea
where $Z_p(g_0, a \mu)$ is a logarithmically divergent
renormalization constant which makes
single insertions of the pseudoscalar density finite,
$\Lambda_n (y)$ denotes a collection of $n$ $q^{(ren)}$ and $\bar q^{(ren)}$
insertions, all at different points:
\be
\Lambda_n (y) \equiv q^{(ren)}(y_1) \dots \bar q^{(ren)}(y_n)
\ee
and $\delta_A^a \Lambda_n (y)$ its axial variation:
\be
\delta_A^a \Lambda_n (y) \equiv \gamma_5 {\lambda^a \over2}
q^{(ren)}(y_1) \dots \bar q^{(ren)}(y_n) + \dots + q^{(ren)}(y_1)
\dots \bar q^{(ren)}(y_n) \gamma_5 {\lambda^a \over2}
\ee

In order to study the $\bar X^a (x)$ insertion in the r.h.s of
eq.(\ref{form}) we start considering:
\begin{equation}
\Phi^{ab} \equiv Z_p
\int d^4 y\langle
\bar O^a_5(x) \ \bar q(0) \gamma_5 {\lambda^b \over 2} q(0) \ \Lambda_n (y)\rangle
\end{equation}

Although $\bar O^a_5(x)$ and $Z_p \ \bar q(0) \gamma_5 {\lambda^b \over 2} q(0)$, being renormalized operators,
have finite insertions with fundamental
fields, $\Phi^{ab}$ is still u.v. divergent
due to short distance non-integrable singularities when $x \rightarrow 0$.
We will consider here, for simplicity, the case $b \neq 0, a$.
With an appropriate choice of $C_s$
we can construct out of $\Phi^{ab}$ a finite quantity as:
\begin{equation}
\Phi^{ab}_R= \Phi^{ab} - Z_p Z_5 {{C_s} \over a} d^{abc}
\langle \bar q(0) {\lambda^c \over 2} q(0) \
\Lambda_n (y) \rangle \label{fir}
\end{equation}

As before we get a restriction on $C_s$ by
applying $\csm$ to both sides of eq.(\ref{fir}):
\begin{eqnarray}
& & \csm \Phi^{ab}_R \equiv \csr \Phi^{ab}_R= \label{venticinque}\\
& & = - (\gamma_{\bar O_5}(g) + \gamma_p(g) + n \gamma_q(g))
\Phi_R^{ab}+ \nonumber \\
& & - {{Z_p Z_5} \over a} d^{abc}
\langle \bar q(0) {\lambda^c \over 2} q(0) \ \Lambda(y) \rangle
\csm C_s\nonumber
\eea
Eq.(\ref{venticinque}) shows that $C_s$ is a function of $g_0$
only\footnote{The dependence on $aM_0$ can be neglected in view of
assumption \ref{erre}.}, so that:
\bea
& & Z_p \int d^4 x\langle
\bar X^a(x) \ \bar q(0) \gamma_5 {\lambda^b \over 2} q(0) \
\Lambda_n(y)\rangle = \label{diciassette} \\
& & ={a \over {Z_5}} \Phi^{ab}(x) \mathop \approx \limits_{a \approx 0}
Z_p C_s(g_0) d^{abc}
\langle \bar q(0) {\lambda^c \over 2} q(0) \ \Lambda_n(y) \rangle \nonumber
\eea
where we exploited the fact that
${a \over {Z_5}} \Phi^{ab}_R \rightarrow 0$, as $a \rightarrow 0$

Eq.(\ref{diciassette}) fully confirms the results of ref.\cite{boc}.
A similar analysis shows that, in general, for $a\neq 0$ and $b=0, \dots, 8$,
we have:
\bea
& & - m_q \int d^4 x\langle \bar q(x)
\gamma_5 \lambda^a q(x) \ P^b(0) \
\Lambda_n (y) \rangle = \label{cwi} \\
& & = d^{abc} \langle S^c(0) \ \Lambda_n(y) \rangle  +
\langle P^b(0) \ \delta_A^a \Lambda_n(y) \rangle \nonumber
\eea
where ($a \neq 0$):
\bea
& & P^0(x) \equiv Z_p (1+ C_p(g_0))
\bar q(x) \gamma_5 {\lambda^0 \over 2} q(x) \label{dens} \\
& & P^a(x) \equiv Z_p \bar q(x) \gamma_5 {\lambda^a \over 2}
q(x) \nonumber \\
& & S^0(x) \equiv Z_p \left[ (1+C_{s^0}(g_0)) \bar q(x)
{\lambda^0 \over 2} q(x) + {1 \over 3}
{{D(g_0,aM_0)} \over a^3} \right] \nonumber \\
& & S^a(x) \equiv Z_p (1+C_s(g_0)) \bar q(x)
{\lambda^a \over 2} q(x) \nonumber
\eea

By evaluating eq.(\ref{cwi}) in the chiral limit, $m_q=0$,
we avoid an u.v. divergence in the l.h.s.\footnote{This divergence
appears in the disconnected component of the Green' function
and is relevant in the definition of the chiral condensate\cite{boc}.},
due to the simultaneous insertion of $\bar q(x)
\gamma_5 \lambda^a q(x)$ and $\bar q(0) \gamma_5 {\lambda^b \over 2} q(0)$
and show that the $P$'s and the $S$'s, defined in eq.(\ref{dens}),
belong to a renormalized $(3,\bar 3) \bigoplus (\bar 3,3)$ representation of
$SU(3) \bigotimes SU(3)$\cite{boc}.

Strictly speaking, the Ward Identity eq.(\ref{cwi}) only provides a
check that the $P^a(x)$'s transform into the $S^a(x)$'s
(for $a=0, \dots, 8$) under an axial transformation.
In principle we should also consider the Ward Identity analogous to
eq.(\ref{cwi}), but with an $S^b(x)$ insertion in the l.h.s. and check
that in the r.h.s. the correct combination $d^{abc}P^c(x)$ appears,
without additional renormalization constants. This consistency is
guaranteed by the fact that Ward identities are obtained by making a
transformation, $\delta_A^a q(x)$ on the quark fields in the functional
integral and the $\delta_A^a$ identically satisfy, on the lattice,
the algebra of $SU(3) \bigotimes SU(3)$:
\bea
& & \delta_A^a \delta_A^b -\delta_A^b \delta_A^a=
-i f^{abc}\delta_V^c \label{var1}\\
& & \delta_V^a \delta_V^b -\delta_V^b \delta_V^a=
-i f^{abc}\delta_V^c \label{var2}
\eea
where $\delta_V^a$ denotes a vector flavor variation.
It is easy to check that eq.(\ref{var1}) provides the required
consistency. An explicit example of this kind of consistency,
in a simpler case, will be given at the end of subsection \ref{42}.

\section{Global Anomalous Symmetries} \label{three}

In this section we will discuss the definition and the renormalization
of the Ward Identities related to the $U(1)$ chiral transformations\cite{old}.

\subsection{Anomalous Ward Identities with Elementary Operator Insertions}

In analogy with the octet case, eq.(\ref{nward}), for a $U(1)$
chiral transformation we have the regularized identity:
\bea
& & \partial^\mu_x \langle A_\mu(x) \ q(y) \ \bar q(z) \rangle=
M_0 \langle \bar q(x) \gamma_5 q(x) \ q(y) \ \bar q(z) \rangle
+ \label{nward0}\\
& & + \delta^{(4)}(x-y) \langle {\gamma_5 \over 2} q(y) \
\bar q(z) \rangle
+ \delta^{(4)}(x-z) \langle q(y) \ \bar q(z) {\gamma_5 \over 2}
\rangle + \nonumber \\
& & + \langle X(x) \ q(y) \ \bar q(z) \rangle \nonumber
\eea
where:
\be
A_\mu(x) \equiv {1 \over 2} \bar q(x) \gamma_\mu \gamma_5 q(x)
\ee
As $X^a(x)$, also $X(x)$ has the form:
\begin{equation}
X(x)=a O_5(x)
\end{equation}
In this case, however, in order to construct a finite operator out
of $O_5(x)$, we have to perform more subtractions.
Instead of eq.(\ref{bar}), we have:
\begin{eqnarray}
& & \bar O_5(x)= Z'_5 \{ O_5(x) + {{\bar M'} \over a} \bar q(x)
\gamma_5 q(x)+ \label{bar0} \\
& & + {(Z'_A -1) \over a} \partial^\mu
A^\mu(x)-{Z_{F\tilde F} \over a} F\tilde F \} \nonumber
\end{eqnarray}
where $F\tilde F$ is any formal regularization of the corresponding classical
operator.

As in the octet case we conclude that $Z'_A$ and $Z_{F\tilde F}$
are finite functions of $g_0$, while:
\begin{equation}
\bar M' = {w'(g_0,aM_0) \over a}
\end{equation}

Renormalizing the quark fields, eq.(\ref{nward0}) becomes:
\bea
& & Z'_A(g_0) \partial^\mu_x \langle A_\mu(x)
\ q^{(ren)}(y) \ \bar q^{(ren)}(z) \rangle= \label{nward00} \\
& & = m'_q \langle \bar q(x) \gamma_5 q(x) \ q^{(ren)}(y) \ \bar q^{(ren)}(z) \rangle
+ Z_{F\tilde F}(g_0) \langle F\tilde F(x) \ q^{(ren)}(y)
\ \bar q^{(ren)}(z) \rangle + \nonumber \\
& & + \delta^{(4)}(x-y)
\langle {\gamma_5 \over 2} q^{(ren)}(y) \ \bar q^{(ren)}(z) \rangle + \delta^{(4)}(x-z)
\langle q^{(ren)}(y) \ \bar q^{(ren)}(z) {\gamma_5 \over 2} \rangle + \nonumber \\
& & + \langle \bar X(x) \ q^{(ren)}(y) \ \bar q^{(ren)}(z) \rangle \nonumber
\eea
where:
\bea
& & m'_q \equiv M_0 -\bar M' \label{chibar0} \\
& & \bar X(x) \equiv {a \over {Z'_5}} \bar O_5(x) \nonumber
\eea
Again, in eq.(\ref{nward00}), the insertion of $\bar X(x)$ vanishes
as $a \rightarrow 0$.
An argument similar to the one used in the octet
case shows the finiteness of the mass insertion\cite{bar}. However
we cannot exclude separate logarithmically divergent contributions from
$Z'_A(g_0) \partial^\mu_x A_\mu(x)$
and $Z_{F\tilde F}(g_0) F\tilde F$.
These divergences are, in fact, present\cite{bar} and require a
logarithmically divergent mixing, in order to define separately finite
operators:
\begin{eqnarray}
& & A_\mu^R(x) \equiv (1- Z_C(g_0,a \mu))
Z'_A(g_0) A_\mu(x) \\
& & F\tilde F^R(x) \equiv
Z_{F\tilde F}(g_0) F\tilde F(x) - Z_C(g_0,a \mu)
Z'_A(g_0) \partial^\mu A_\mu(x) \nonumber
\end{eqnarray}
so that the renormalized anomalous Ward Identity with quark field insertions
becomes:
\bea
& & \partial^\mu_x \langle A_\mu^R(x) \ q^{(ren)}(y) \ \bar q^{(ren)}(z)
\rangle = \label{axial} \\
& & = m'_q \langle \bar q(x) \gamma_5 q(x) \ q^{(ren)}(y)
\ \bar q^{(ren)}(z) \rangle
+ \langle F\tilde F^R(x) \ q^{(ren)}(y) \ \bar q^{(ren)}(z) \rangle + \nonumber \\
& & + \delta^{(4)}(x-y) \langle {\gamma_5 \over 2} q^{(ren)}(y) \
\bar q^{(ren)}(z) \rangle
+ \delta^{(4)}(x-z) \langle q^{(ren)}(y) \ \bar q^{(ren)}(z) {\gamma_5 \over 2}
\rangle \nonumber
\eea
In the anomalous case the normalization of the axial current is, of course,
not fixed by Ward Identities, but must be adjusted according to
some arbitrary prescription.

\subsection{Anomalous Ward Identities with Composite Operator
Insertions} \label{42}

In the presence of a composite operator, as, for instance,
$P^a (y)$, defined in eq.(\ref{dens}), we can write
the regularized and integrated $U(1)$ Ward
Identity\footnote{We choose $a \neq 0$ in order to avoid
disconnected contributions.}:
\bea
& & - \int d^4 x \ \langle [ m'_q \bar q(x)
\gamma_5 q(x) + F\tilde F^R(x) ] \ P^a(y) \ \Lambda_n (z)
\rangle = \label{form0i} \\
& & = \langle P^a (y) \ \delta_A \Lambda_n (z) \rangle +
Z_p \ \langle \bar q(y){\lambda^a \over 2}
q(y) \ \Lambda_n (z) \rangle \nonumber \\
& & + \int d^4 x \ \langle \bar X (x)
\ P^a (y) \ \Lambda_n (z)\rangle \nonumber
\eea
with an obvious meaning for $\delta_A$.
For the insertion of $\bar X(x)$ in eq.(\ref{form0i}),
we can construct an argument
completely parallel to the one leading to eq.(\ref{diciassette}),
which gives:
\bea
& & \int d^4 x\langle
\bar X(x) \ P^a (y) \
\Lambda_n(z) \rangle \approx \label{ins} \\
& &\mathop \approx \limits_{a \approx 0}
Z_p C'(g_0)
\langle \bar q(y) {\lambda^a \over 2} q(y) \ \Lambda_n(z) \rangle \nonumber
\eea
so that eq.(\ref{form0i}) becomes:
\bea
& & - \int d^4 x \langle [ m'_q \bar q(x)
\gamma_5 q(x) +
F\tilde F^R(x) ] \ P^a (y) \ \Lambda_n(z) \rangle =  \label{wi0} \\
& & = \langle P^a (y) \ \delta_A \Lambda_n (z) \rangle +
Z_p (1+C'(g_0)) \langle \bar q(y) {\lambda^a \over 2} q(y) \
\Lambda_n(z) \rangle \equiv \nonumber \\
& & \equiv \langle P^a (y) \ \delta_A \Lambda_n (z) \rangle +
{1+C'(g_0) \over 1+C_s(g_0)} \langle S^a(y) \
\Lambda_n(z) \rangle \nonumber
\eea
Through operator product expansion one can convince oneself that
the double insertion
of composite operators in eq.(\ref{wi0}) is integrable without
further subtractions. In fact in the product of
$m'_q \ \bar q(x) \gamma_5 q(x)$ and $P^a (y)$ at $x \approx y$, the identity
operator is missing because of the flavour structure ($a\neq 0$)
and the next contributing operator has
dimension $3$, giving rise to an integrable singularity. A similar
argument holds for the double insertion of $F\tilde F^R(x)$
and $P^a (y)$.

Eq.(\ref{wi0}) is the prototype of the $U(1)$ integrated chiral
Ward identities with composite operator insertions and can be used
to solve the following consistency problem.

When sandwiched between on-shell states, $\hat A^a_\mu (x)$
and $A^R_\mu (x)$ satisfy the partial conservation equations:
\bea
& & \partial^\mu \hat A^a_\mu (x)= 
m_q \bar q(x) \lambda^a \gamma_5 q(x) \label{parto} \\
& & \partial^\mu A^R_\mu (x)= 
m'_q \bar q(x) \gamma_5 q(x)+ F\tilde F^R(x) \label{parts}
\eea

While in eq.(\ref{parts}) the mass insertion is proportional to
$m'_q \equiv M_0 - \bar M'$, in eq.(\ref{parto}) it
is proportional to $m_q \equiv M_0-\bar M$. Formally,
in the continuum, the mass insertions in eqs.(\ref{parto}) and
(\ref{parts}) have the same coefficient, the quark mass, 
so that they vanish together in the chiral limit. 

We want to show that this is also true for the renormalized
eqs.(\ref{parto}) and (\ref{parts}):
$m_q$ and $m'_q$ are in fact proportional through a finite coefficient,
so that both vanish in the chiral limit. 

We integrate eq.(\ref{wi0}) over $y$:
\bea
& & - \int d^4 x d^4 y \ \langle [ m'_q \bar q(x)
\gamma_5 q(x) +
F\tilde F^R(x) ] \ P^a (y) \ \Lambda_n(z) \rangle =  \label{wi0i} \\
& & = \int d^4 y \ \langle P^a (y) \ \delta_A \Lambda_n (z) \rangle +
Z_p (1+C'(g_0)) \int d^4 y \langle \bar q(y) {\lambda^a \over 2}
q(y) \ \Lambda_n(z) \rangle \nonumber
\eea
and consider the chain of equalities:
\bea
& & - \int d^4 x d^4 y 
\langle m_q \bar q(y)
\lambda^a \gamma_5 q(y) [ m'_q
\bar q(x)
\gamma_5 q(x) + F\tilde F^R(x) ] \Lambda_n(z) \rangle= \label{alfa}\\
& & = m'_q {{1+C_s(g_0)} \over {1+C_p(g_0)}} \int d^4 x
\langle \bar q(x) \lambda^a q(x) \ \Lambda_n(z)
\rangle + \label{beta} \\
& & + \int d^4 x \langle \left[ m'_q \bar q(x)
\gamma_5 q(x) + F\tilde F^R(x) \right]
\delta_A^a \Lambda_n(z) \rangle = \nonumber \\
& & = m_q (1+C'(g_0)) \int d^4 y
\langle \bar q(y)
\lambda^a q(y) \ \Lambda_n(z) \rangle + \label{gamma} \\
& & + \int d^4 y 
\langle m_q \bar q(y)
\lambda^a \gamma_5 q(y) \ \delta_A \Lambda_n(z) \rangle \nonumber
\eea
Eq.(\ref{beta}) follows from eq.(\ref{alfa}) through the octet Ward
Identity, eq.(\ref{cwi}), and the octet chiral invariance
of $\int d^4 x F \tilde F^R(x)$\footnote{$\int d^4 x F \tilde F^R(x)$
only depends on gluon fields.}, while eq.(\ref{gamma})
follows from eq.(\ref{alfa}) through eq.(\ref{wi0i}).

The terms containing $\delta_A^a \Lambda_n(z)$ and
$\delta_A \Lambda_n(z)$, in eqs.(\ref{beta}) and (\ref{gamma}), can be
further transformed.
The integrated octet Ward Identity, eq.(\ref{renc}), in fact, gives:
\be
m_q \int d^4 y \langle \bar q(y)
\lambda^a \gamma_5 q(y) \ \delta \Lambda_n(z) \rangle =
- \langle \delta_A^a \delta_A \Lambda_n(z) \rangle
\ee
while, integrating the $U(1)$ Ward Identity eq.(\ref{axial}), we get:
\be
\int d^4 x \langle [ m'_q \bar q(x)
\gamma_5 q(x) + F\tilde F^R(x)]
\delta^a \Lambda_n(z) \rangle
= - \langle \delta_A \delta_A^a \Lambda_n(z) \rangle
\ee
Since the octet and the $U(1)$ axial transformations commute:
\be \delta_A \delta_A^a = \delta_A^a \delta_A \label{comm}
\ee
eqs.(\ref{beta}) and (\ref{gamma}) imply:
\be
m'_q {{1+C_s(g_0)} \over {1+C_p(g_0)}} = m_q (1+C'(g_0))
\ee
which proves the required consistency.

We can further show that:
\be
C'(g_0)=C_s(g_0) \label{equ}
\ee
In fact eq.(\ref{wi0}) tells us that\footnote{We still consider
$a\neq 0$, for simplicity.}:
\bea
& & P^a(x) \\
& & {1+C'(g_0) \over 1+C_s(g_0)} S^a(x) \nonumber
\eea
belong to an irreducible, renormalized representation of the chiral $U(1)$
group. From the commutativity of $U(1)$ and octet
transformations ($b \neq 0, a$), eq.(\ref{comm}), we have:
\bea
& & \delta \delta ^b P^a(x)={1+C_s(g_0) \over 1+C'(g_0)}
d^{bac} P^c(x) =\\
& & =\delta ^b \delta P^a(x) ={1+C'(g_0) \over 1+C_s(g_0)}
d^{bac} P^c(x) \nonumber
\eea
so that:
\be
{1+C_s(g_0) \over 1+C'(g_0)}=\pm 1
\ee
where the sign ambiguity can be reabsorbed into a redefinition of
the abelian $U(1)$ charge, thus proving eq.(\ref{equ}).

\section*{Acknowledgements}

I thank the CERN Theory Division for the kind hospitality.

\noindent I also thank Giancarlo Rossi for many discussions
and suggestions and
Laurent Lellouch for pointing out a mistake in an earlier version of the
manuscript.

\end{document}